\documentclass[preprint,eqsecnum,aps]{revtex4}
\usepackage{amsmath}

\begin{document}

\title{Rotation and Spin in Physics\footnote{This paper is dedicated
to the memory of Professor J. A. Wheeler}}

\author{R. F. O'Connell}
\affiliation{Department of Physics and Astronomy, Louisiana State
University, Baton Rouge, LA 70803-4001}
\date{\today}

\begin{abstract}
We delineate the role of rotation and spin in physics, discussing in order
Newtonian classical physics, special relativity, quantum mechanics,
quantum electrodynamics and general relativity.  In the latter case, we discuss
the generalization of the Kepler formula to post-Newtonian order $(c^{-2}$)
including spin effects and two-body effects.  Experiments which verify the
theoretical results for general relativistic spin-orbit effects are discussed
as well as efforts being made to verify the spin-spin effects.
\end{abstract}

\maketitle

\section{Introduction}

This article is essentially a companion to an article which I recently
prepared for the 2007 Varenna summer school \cite{oconnell07}.  The latter
deals in detail with spin effects in general relativity at the
post-Newtonian level (order $c^{-2}$) with particular emphasis on the
theory, and related experiments, of the two-body Kepler problem extended to
include spin.  Here, we use "spin" in the generic sense of meaning
"internal spin" in the case of an elementary particle and "rotation" in
the case of a macroscopic body.

My work with Barker on the general relativistic theory
\cite{barker70,barker75,barker76} brought to the fore some interesting conceptual
matters dealing with the relation between velocity and momentum
(especially for particles with spin), the non-uniqueness of spin
supplementary conditions and the choice of coordinates even at the
classical special relativistic level, the fact that a spinning particle
necessarily has a minimum radius, the corresponding concepts in quantum
theory (relating to such topics as the Foldy-Wouthuysen transformation and
position operators) and the fact that the spin effects in quantum
electrodynamics (obtained from one-photon exchange) have their analogy in
general relativity (obtained from one-graviton exchange or purely
classical calculations) to the extent that except for (important)
numerical factors, the latter results may be obtained from the former by
simply letting $e^{2}\rightarrow Gm_{1}m_{2}$.

Thus, we are motivated to present in a systematic way the role of
rotation and spin in physics generally.  So, we discuss, in order,
Newtonian classical physics, special relativity, quantum mechanics,
quantum electrodynamics and general relativity, in sections 2 to 6,
respectively.  Our conclusions are presented in Section 7. The closest discussion
of this nature in the literature is the book by Corben \cite{corben68}.  However,
whereas his list of references proved a useful resource, his emphasis is on
elementary particles and he does not consider general relativity, which is our
emphasis.

\section{Newtonian classical physics}

Newton's second law is generally written in the form

\begin{equation}
\vec{F}=m\frac{d\vec{v}}{dt}, \label{rs21}
\end{equation}
where $\vec{v}$ is the velocity.  Next, the momentum $\vec{p}$ of a
particle is essentially defined as $m\vec{v}$ so that we have

\begin{equation}
\vec{F}=\frac{d\vec{p}}{dt}. \label{rs22}
\end{equation}

As French has emphasized \cite{french71}, Newton in the \emph{Principia}
did not write (\ref{rs21}) but in essence, wrote (\ref{rs22}).  Also, only
in certain frames of reference, so-called inertial frames, are Newton's
laws valid.  All frames, moving uniformly in a straight line relative to a
particular inertial frame, constitute an infinity of inertial frames
where "--the properties of space and time are the same, and the laws of
mechanics are the same--" \cite{landau76}, and such "--frame(s) of
reference can always be chosen in which space is homogenous and isotropic
and time is homogeneous" \cite{landau76}.  This leads us to a new concept 
which assigns a broader independent meaning to momentum.  It follows
from the introduction of the Lagrangian $L(q,\dot{q},t)$ of a system,
where $q$ refers to position and $\dot{q}=(dq/dt)$ is the velocity.  As a
result, the associated Euler-Lagrange equations for a closed system lead to
conservation laws of energy (resulting from the homogeneity of time), momentum
(from the homogeneity of space) and angular momentum (from the isotropy of
space) where, in particular, the momentum is referred to as the canonical
momentum $\vec{P}$, as distinct from the mechanical momentum $\vec{p}=m\vec{v}$,
where

\begin{equation}
\vec{P}=\frac{\partial L}{\partial\vec{v}}. \label{rs23}
\end{equation}
In general, $\vec{P}$ and $\vec{p}$ are not equal, an example being the case of a
particle with charge $q$ in an electromagnetic field, where the force is velocity
dependent, in which case

\begin{equation}
\vec{P}=\vec{p}+\frac{q\vec{A}}{c}. \label{rs24}
\end{equation}
Thus, whereas $\vec{P}$ is conserved, $\vec{p}$ is not conserved in
general.  Moreover, in the passage to quantum mechanics, it is the
canonical variables which are of importance in forming operators.  In
addition, it is $\vec{P}$ that enters into the conserved angular momentum
$\vec{L}=\vec{r}\times\vec{P}$. Of course, in this case, the reason why
$\vec{P}$ and $\vec{v}$ are not proportional is due to the presence of an
external field.  However, in special relativity, we shall see that, even for a
free particle, even $\vec{p}$ and $m\vec{v}$ are, in general, not equal to each
other.

Turning to the rotation of macroscopic bodies, we define $\vec{S}$ as the
spin or intrinsic angular momentum.  In general,

\begin{equation}
S_{i}=I_{ij}\omega_{j}~~~~~(i,j=1,2,3), \label{rs25}
\end{equation}
where $\vec{\omega}$ is the angular velocity and $I_{ij}$ are the
components of the inertia tensor.

Based on observations, the earth is an inertial frame if one neglects its
rotation.  However, taking the "fixed stars" as an inertial system with
"space axes", the earth rotates with an angular velocity $\vec{\omega}$
relative to these axis, giving rise to Coriolis and centrifugal forces.
Thus, following Goldstein \cite{goldstein92} and writing

\begin{equation}
\vec{v}_{s}=\vec{v}_{v}+\vec{\omega}\times\vec{r}, \label{rs26}
\end{equation}
where $\vec{v}_{s}$ and $\vec{v}_{r}$ are the velocities of a particle
relative to the space and the earth's rotating set of axis, the equation

\begin{equation}
\vec{F}_{s}=ma_{s}, \label{rs27}
\end{equation}
in the space system translates to

\begin{equation}
\vec{F}_{r}=ma_{r}=ma_{s}-2m(\vec{\omega}\times\vec{v}_{r})-m\vec{\omega}
\times(\vec{\omega}\times\vec{r}), \label{rs28}
\end{equation}
in the rotating system.  The second and third terms are the familiar
Coriolis and centrifigal forces, respectively.  Of particular interest is a
result, not found in \cite{goldstein92} but derived in \cite{landau76}, that
the momentum of a particle is the same in both frames of reference so that, in
the rotating system,

\begin{equation}
\vec{p}=m\vec{v}+m(\vec{\omega}\times\vec{r}). \label{rs29}
\end{equation}
Thus, here we have a case where the momentum is different from $m\vec{v}$
purely due to the non-inertial nature of the rotating system.  We also
remark that the angular momenta are also equal in both systems but the
energy in the rotating system $E_{r}=E_{s}-\vec{L}\cdot\vec{\omega}$
\cite{landau76}.

Finally, we remark that the earth is not a perfect sphere but is
in fact an "asymmetric top", so that $\vec{\omega}$, the angular velocity
of the earth, as distinct from $\vec{S}$ does not remain fixed in space
but, instead, it executes what is known as polhode motion
\cite{landau76,goldstein92}.  This force-free precession of the earth's
axis is a key to the explanation of Chandler's precession of the earth's axis and
is also proving to be a major bugbear in the analysis of the GP-B gyroscope
experiment \cite{brumfiel06,stanford06}.

\section{Special Relativity}

Just as the angular momentum $\vec{L}$ generalizes to an anti-symmetric
second rank tensor so does the 3-vector $\vec{S}$ generalizes to $S_{\alpha
\beta}=-S_{\beta\alpha}$.  Another possibility is to define an axial 4-vector
$S^{\mu}$ which reduces to the 3-vector $\vec{S}$ in the rest-frame of
the particle. In fact, this can be achieved by defining

\begin{equation}
S_{\alpha}\equiv\frac{1}{2}\epsilon_{\alpha\beta\sigma\tau}\quad
S^{\beta\sigma}U^{\tau}, \label{rs31}
\end{equation}
where $\epsilon_{\alpha\beta\sigma\tau}$ is the completely antisymmetic
Levy-Civita tensor, $U^{\tau}=(\gamma ,\vec{v}/c)$ is the familiar 4-velocity,
and $S^{\alpha}=(0,\vec{S})$ in the rest frame where $\vec{v}=0$.  Hence, using
the fact that $\epsilon_{\alpha\beta\sigma\tau}$ is antisymmetric in $\alpha$
and $\tau$, we obtain

\begin{equation}
U^{\alpha}S_{\alpha}=0, \label{rs32}
\end{equation}
so that the 4-vectors $U^{\alpha}$ and $S_{\alpha}$ are not only orthogonal in
the rest frame as constructed but are also orthogonal in all frames.  In
addition, using the properties of the Levi-Citiva symbol (Ref. \cite{corben68},
p. 79) and
$U^{\alpha}U_{\alpha}=-1$, (\ref{rs31}) may be inverted to give

\begin{equation}
S^{\alpha\beta}=\epsilon^{\alpha\beta\sigma\tau}S_{\sigma}U^{\tau}.
\label{rs33}
\end{equation}
Eq. (\ref{rs32}) is a spin supplementary condition (SSC) which ensures that
even when $\vec{v}$ is non-zero, $S^{\mu}$ has only three independent
components and similarly for $S^{\alpha\beta}$.  For a \underline{free}
particle (no external forces or torques)

\begin{equation}
\frac{dS^{\mu}}{dt}=0, \label{rs34}
\end{equation}
but, nevertheless, as detailed in \cite{corben68},the particle moves in a
circle in a plane normal to $\vec{S}$ with a radius

\begin{equation}
\vec{r}=-\frac{\vec{v}\times\vec{S}}{mc^{2}}. \label{rs35}
\end{equation}
Such a motion is reminiscent of Zitterbewegung in relativistic quantum
mechanics \cite{sakurai67}, as we shall discuss below.  We will also see
how such a quantity as $\vec{r}$ appears in the discussion of spin
effects on orbital motion in post-Newtonian general relativistic theory. 
For now, we note that such a problem can be circumvented by choosing a
different SSC based on the fact that there are essentially two basic rest
systems for the particle, corresponding to either $\vec{v}=0$ or
$\vec{p}=0$.  This is also connected to the fact that, as Moller has shown
\cite{moller49,moller72}, a spinning body has a minimum radius equal to
$|\vec{S}|/mc$. As a result, the definition of a rest frame is related to a
choice of SSC which, in turn, is related to the choice of a center-of-mass for
the spinning particle \cite{barker74}.  Thus, switching from one SSC (or
rest-frame) to another is exactly the same as shifting the center of mass by a
Lorentz transformation.

Turning to the case where a particle is acted on by a force $f^{\mu}$ without
experiencing any torque, Weinberg \cite{weinberg72} has shown that

\begin{equation}
\frac{dS^{\alpha}}{d\tau}=\left(S_{\beta}\frac{f^{\beta}}{m}\right)U^{\alpha},
\label{rs36}
\end{equation}
which leads to the famous Thomas precession \cite{thomas26,jackson98}.  The
usual discussion of Thomas precession is concerned with precession of a
3-vector, leading to a reduction by $\frac{1}{2}$ in the spin-orbit energy of an
atomic electron (taking the gyromagnetic ratio $g$ to be 2).  This analysis was
generalized by Bargmann et al. \cite{bargmann59} to obtain $dS^{\alpha}/d\tau$
in its most simple form by assuming from the outset that the momentum and
velocity of a spinning particle are proportional \cite{corben68}.  Thus, they
were led to the so-called BMT
equation for the classical relativistic equation of
motion for spin in uniform or slowly varying external fields and its
various applications.  Details of this work are given in
Jackson's well-known book \cite{jackson98}, along with the derivation of Thomas's
equation of motion of the spin vector from which Thomas precession emerges. 
Important applications of the BMT equation are listed in \cite{corben68}.

\section{Quantum Mechanics}

As detailed in all the textbooks, spin was introduced into non-relativistic
quantum mechanics by Goudsmit and Uhlenbeck \cite{uhlenbeck26}.  It should also
be mentioned that the Wigner distribution, which represents an alternative
formulation of quantum mechanics \cite{hillery84,oconnell} has also been
extended to incorporate spin \cite{oconnell84}.

Relativistic quantum mechanics was initiated by Dirac's equation for the
electron.  However, the Dirac equation for a \underline{free} particle, leads
to the conclusion that the momentum and velocity are not simply related
\cite{sakurai67}.  Thus, whereas the plane-wave solutions  are eigenfunctions of
the momentum operator, the velocity operator is not a constant of the motion
with the implication that the free electron not only displays uniform
rectilinear motion but also executes very rapid oscillations, called
Zitterbewegung by Schrodinger who, as early as 1930, first discussed such
motion.  However, since we noted earlier that all classical spinning bodies
have a minimum radius $|\vec{S}|/mc$, it is reasonable to assume that the
electron (for which $|\vec{S}|=\hbar/2$) cannot be localized to a size smaller
than $\lambdabar_{c}/2$, where $\lambdabar=\hbar/mc$ is the Compton
wavelength.  Hence, analogous to the Lorentz transformations which shift the
center of mass of a classical particle with spin, for an electron we can change
the coordinate operator by means of a unitary Foldy-Wouthuysen transformation
\cite{foldy50,rose61}.  In this representation the velocity operator is
proportional to the momentum operator but, as emphasized by Sakurai (p. 177 of
\cite{sakurai67}), "nonlocality --- is the price we must pay."  By contrast,
"--a well-localized state contains, in general, plane-wave components of
negative energy" \cite{sakurai67}.  In earlier work and in a similar vein,
Newton and Wigner \cite{newton49} created the most localized state possible for
a free particle without spin, and also the corresponding position operator, by
using only plane-wave components of positive energies (solutions of the
Klein-Gordon equation) and concluded that it is not a delta function as in the
nonrelativistic case; instead, the wave function goes as $r^{-5/2}$ for small $r$
and falls off exponentially for large $r$ and, in general its characteristic
extension in space is $\approx\lambdabar_{c}$.  This work was then extended to
particles with spin for which position operators and quantum states were also
obtained.  The results obtained are unique subject to the restriction that the
orbital angular momentum of the localized state has the value
$\ell=0$.  However, for spin 1/2 particles, it was shown by the present author
and Wigner \cite{oconnell78} that another choice of position operator is
possible, given by

\begin{equation}
\vec{Q}=\vec{q}+(\vec{p}\times\vec{\sigma})/p^{2}, \label{rs41}
\end{equation}
where $\vec{q}$ is the original position operator, $\vec{Q}$ is the new position
operator, and where the components of $\vec{\sigma}$ denote the Pauli spin
matrices.  In addition, it was shown \cite{oconnell77} that

\begin{equation}
\left<\vec{q}(t)\right>=\left<\vec{q}(0)\right>+t\left<\vec{v}\right>,
\label{rs42}
\end{equation}
where here the velocity operator is $\vec{v}=(c^{2}\vec{p}/E)$, where $E$ is
the total energy, including the rest mass.  Thus, the movement of the mean
position of a free particle obeys the classical Newtonian equation except for
$m\rightarrow (E/c^{2})$.  Moreover, a similar result was obtained
\cite{oconnell78} for $\left<\vec{Q}(t)\right>$.  Even for the case of a spin
zero particle, (\ref{rs42}) is a non-trivial result because, as emphasized in
\cite{oconnell77}, "--the situation in quantum mechanics is not immediately
obvious because although $\vec{p}$ has a natural definition - it is the
generator of spatial translations of the state vector - this is not so for
$\vec{v}$ - the "velocity" does depend on the definition of the position--". 
Finally, we note that the choice of state vectors given by (\ref{rs41}) and
(\ref{rs42}) leads to motion free from Zitterbewegung, just as we saw  for the
Foldy-Wouthuysen position operator.

\section{Quantum Electrodynamics (QED)}

The initial (semi-classical) treatment of particles and atoms with the
electromagnetic field considered the latter to be classical.  This was
dramatically changed in 1927 when Dirac quantized the electromagnetic field
\cite{dirac27}.  Thus, for fields in general, the emphasis then switched from
trying to construct relativistic potentials to the consideration of suitable
Lagrangians and the use of Feynman diagrams, $S$-matrix theory
\cite{wheeler37} and other techniques.  These are particularly useful in
carrying out perturbation calculations.  In contrast to nonrelativistic quantum
theory (where potentials are used) in relativistic field theory, interactions
are considered to arise from the exchange of quanta.  Thus, in QED, the quanta
are photons of spin 1 (whereas, as we shall discuss at length in the next
section, gravitational interactions can be considered to be mediated by spin 2
gravitons).  In essence, each diagram gives rise to a covariant matrix element
whose Fourier transform "--enables us to construct an effective
three-dimensional potential (to be used in connection with the Schrodinger
equation)--" (Ref. \cite{sakurai67}, p. 259).  A detailed description of this
technique is given in \cite{berestetskii71}; in particular, they consider the
electromagnetic interaction of two particles with masses $m$, and $m_{2}$ (such
as an electron and a muon) for which they first calculate the particle
interaction operator in the momentum representation.  Next, by taking Fourier
transforms, they obtain the corresponding operator in the coordinate
representation (Ref.
\cite{berestetskii71}, p. 283, Eq. (83.15)); the latter contains terms of
purely orbital origin as well as spin-orbit and spin-spin terms, and
Darwin-like terms. As we shall see in the next section, an analogous method for
the gravitational interaction of two spinning bodies leads to similar type terms.

\section{General Relativity}

For a free particle with spin $S^{\alpha}$, in special relativity we had the
result given by (\ref{rs34}).  The covariant generalization of this formula is
\cite{weinberg72}

\begin{equation}
\frac{DS^{\mu}}{D\tau}=0, \label{rs61}
\end{equation}
where $D$ denotes covariant differentiation.  Also, in the presence of a
gravitational field, (\ref{rs36}) generalizes to \cite{weinberg72,ciufolini95}

\begin{equation}
\frac{DS^{\mu}}{D\tau}=S_{\nu}\frac{DU^{\nu}}{D\tau}U^{\mu}\equiv
S_{\nu}a^{\nu}U^{\mu}, \label{rs62}
\end{equation}
where $a^{u}=\frac{f^{\mu}}{m}$ is the 4-acceleration.  This is referred to as
Fermi-Walker transport of the spin vector in its motion along a curve
$x^{\alpha}$ with tangent vector $U^{\alpha}$.  In the special case where
$f^{\mu}=0$ (free fall), the right-side of (\ref{rs62}) is zero and the
Fermi-Walker transport reduces to parallel transport along the geodesic, given
by (\ref{rs61}), which may be written explicitly as

\begin{equation}
\frac{dS_{\mu}}{d\tau}=\Gamma^{\lambda}~_{\mu
\nu}S_{\lambda}\frac{dx^{\nu}}{d\tau},
\label{rs63}
\end{equation}
where $\Gamma^{\lambda}_{\mu \nu}$ is the Christoffel symbol.  Next, Papapetrou
\cite{papapetrou51,corinaldesi51} considered the motion of a small mass with spin in a
gravitational field in order to obtain the deviation from geodesic motion. 
This deviation is expressed in both orbital and spin equations of motion, given
by

\begin{equation}
\frac{D}{Ds}\left(mu^{\alpha}+u_{\beta}\frac{DS^{\alpha\beta}}{Ds}\right)+
\frac{1}{2}S^{\mu v}u^{\sigma}R^{\alpha}_{\nu\sigma\mu}=0, \label{rs64}
\end{equation}
and

\begin{equation}
\frac{DS^{\alpha\beta}}{Ds}+u^{\alpha}u_{\rho}~\frac{DS^{\beta\rho}}{Ds}-
u^{\beta}u_{\rho}~\frac{DS^{\alpha\rho}}{Ds}=0, \label{rs65}
\end{equation}
respectively, where $R^{\alpha}_{\nu\sigma\mu}$ is the Riemann curvature tensor
and $S^{\alpha\beta}$ is given by (\ref{rs33}).  This was the starting-point
used by Schiff \cite{schiff64} in his derivation of the theory underlying the
GP-B experiment \cite{stanford06}, which involved calculations to
post-Newtonian order.  However, Barker and the present author \cite{barker70}
applied a completely different approach to the problem than the method used by
Papapetrou and Schiff, obtaining results which agreed with those of Schiff for
the spin precession but apparently different for the orbital equations of
motion.

The method used in \cite{barker70} was based on a quantum approach to
gravitation developed by Gupta \cite{gupta57} and his collaborators
\cite{barker66}. In particular, the gravitational interaction between two spin
1/2 Dirac particles was calculated by means of the exchange of a spin 2
graviton, analogous to what we discussed in Section 5 in the case of the
electromagnetic interaction of two charged spin 1/2 Dirac particles.  In fact,
comparing the results in the latter case for the interaction of, say, an
electron and muon (Ref. \cite{berestetskii71}, Eq. (83.15)) with these obtained
in the former case \cite{barker70}, we were able to conclude that all such
effects in QED have their analogy in general relativity and, except for
(important) numerical factors, the latter results may be obtained from the
former by simply letting $e^{2}\rightarrow Gm_{1}m_{2}$.  Based on the
universality of gravitational interactions, the transition from the
microscopic to the macroscopic is achieved by the simple replacement
$\frac{1}{2}~\hbar~\vec{\sigma}\rightarrow\vec{S}$ \cite{barker70}.  In a later
publication
\cite{barker742}, we traced the apparent difference betweeen our results and
those of Schiff for the orbital equations of motion to the use of two different
coordinate vectors, reflecting the choice of two different SSC's which in turn
imply different choices for the center-of-mass of the gyroscope and also imply
different relationships between the momentum and velocity vectors.  Thus, as
outlined in
\cite{barker742}, Corinaldesi-Papapetrou and Schiff used $S^{i0}=0$ as their
SSC; Pirani used
$S^{\alpha\beta}U_{\beta}=0$ whereas Moller, Tulczyjew and Dixon used
$S^{\alpha\beta}P_{\beta}=0$.  In particular, the difference in the coordinate
vectors used by all investigators is always of the order
$(\vec{v}\times\vec{s})/mc^{2}$, the same quantity which arose in the discussion
of anomalous motion in special relativity [see (\ref{rs35})] and not unrelated
to the choice of position operators in relativistic quantum theory [see
(\ref{rs41})].  However, regardless of the choice of SSC, in all cases identical
results are obtained for orbital precessions.  We also remark that, in contrast
to the orbital case, the form of the spin equations of motion is the same for all
choices of SSC's; the reason is that spin contributions are always order
$c^{-2}$ compared to the lowest-order Newtonian term appearing in the equation
of motion.

Turning to experimental tests, Schiff's 1960 paper stimulated the birth of the
GP-B experiment.  However, because results are only now emerging, the initial
euphoria which this experiment generated has waned \cite{lawler03} due to the
fact that other investigators have verified the existence of spin-orbit effects
in gravitation but, at least, the GP-B initial report \cite{stanford06} claims
increased accuracy for the results obtained for the same quantity.  The other
investigations (which are discussed in more detail in \cite{oconnell07}, where
the emphasis is on the Kepler problem and its post-Newtonian generalization,
including spin) include (a) lunar-laser-ranging measurements of the lunar
perigee \cite{williams96,williams04} based on the fact that the earth-moon
system is essentially a gyroscope in the field of the sun \cite{oconnell05};
(b) the Ciufolini Lageos experiment \cite{ciufolini04}, based on the
determination of earth-satellite distances with a precision of a few $mm$; (c)
the binary pulsar PSRB1534+12 observations \cite{stairs04} based on a
determination of time evolutions involving the spin direction.

The advantage of these experiments is that they are on-going (in contrast to
the GP-B experiment) and thus improved accruacy is likely to be achieved.  In
particular, Ciufolini \emph{et al.}, whose \emph{Nature} paper \cite{ciufolini04} is the first very credible demonstration of gravitomagnetism, are making use of the continual improvement in
accuracy of earth gravity models \cite{ciufolini06}.

The observations in \cite{stairs04} while limited in precession also provide an
initial test of a 2-body result for the spin-orbit precession
\cite{barker75,barker76}.  What we found is that, for the spin precession of
body 1, the replacement in the one-body formula is $m_{2}\rightarrow m_{2}+(\mu
/3)$ and, in the case of the spin precession of body 2, the replacement is
$m_{1}\rightarrow m_{1} +(\mu /3)$, where $\mu$ is the reduced mass.  This result
has now been verified by at least 7 other authors, the simplest being a
calculation which makes use of the one-body result, followed by a transformation
to the center-of-mass of the two-body system \cite{chan77}.  A potentially
more promising candidate is the double binary system \cite{kramer06} for which
spin precession angles of $4.8^{o}$/yr (pulsar A) and $5.1^{o}$/yr (pulsar B)
have been predicted, based on our theoretical results \cite{barker75,barker76}.

Finally, we remark on what we consider to be the most theoretically appealing
alternative to Einstein's theory in which the spin of matter, as well as its
mass, plays a dynamical role.  This idea was initiated by Cartan
\cite{cartan24} and extended by Sciama \cite{sciama64} and Kibble
\cite{kibble61} using gauge theory.  Utiyama initially investigated Einstein's
theory and the Cartan theory within the framework of the homogeneous and
inhomogeneous (Poincar\'{e}) groups, respectively \cite{utiyama56}. For a
review, we refer to \cite{hehl76}.  We investigated possible additional spin
interactions based on this so-called torsion theory and we found that spin-spin
contact interactions, additional to these associated with Einstein's theory,
occur \cite{oconnell772,oconnell773}.  However, as it turns out, they do not
contribute to the macroscopic experiments presently being carried out.

\section{Conclusions}

In surveying the role of rotation and spin over the broad range of physics,
several important concepts were considered.  In non-relativistic classical
physics, we recalled that Newton's second law only holds in an inertial system
(in which case the mechanical momentum $\vec{p}=m\vec{v}$) but does not hold
in a rotating system which is non-inertial.  In the latter case, $\vec{p}$ is
not proportional to $\vec{v}$.  Also, the introduction of a Lagrangian $L$
for a system brings into consideration a new type of momentum, the canonical
momentum $\vec{P}$, which is the derivative of $L$ with respect to $\vec{v}$
and which, in the case of external fields, is different than $\vec{p}$.  The
intrinsic angular momentum (spin) $\vec{S}$ also finds a natural definition
within the Newtonian framework.  However, in special relativity, one must treat
spin as either a 4-vector $S^{\mu}$ or else as a second rank antisymmetric
tensor and define a relationship between them which requires the introduction of
a spin supplementary condition (SSC).  This led to the realization that there
are essentially two basic rest systems, corresponding to either $\vec{v}=0$
or $\vec{p}=0$, with the possibility of considering a variety of choices for a
SSC instead of the choice given in (\ref{rs32}).  As a result, different
results for $\vec{p}$ in terms of $\vec{v}$ emerge and, concomitantly, this
also implies that the position vector of even a microscopic spinning particle
is not an observable in special relativity.

Next, turning to quantum mechanics, we encountered similar phenomena.  In
particular, Zitterbewegung was found to be a feature of a particular choice of
coordinate operator associated with Dirac's formulation of relativistic
electron theory.  It could be removed by carrying out a unitary transformation
to a different coordinate for which Zitterbewegung disappears and simple
proportionality between $\left<\vec{p}\right>$ and $\left<\vec{v}\right>$
holds.  However, even for a free particle with or without spin, it was found
necessary to choose a very specific position operator if one uses only
plane-wave components of positive energy, in which case the particle is
non-local and has a characteristic extension $\approx\lambdabar_{c}$.  Extending
consideration to the QED domain, we saw that the modus operandi is to consider
the interactions as arising from the exchange of spin1 photons, leading to
spin-orbit, spin-spin and other terms.  A further extension of these ideas to
the realm of general relativity called for the exchange of spin 2 gravitons. 
This led to a systematic derivation of spin-orbit and spin-spin terms, to order
$c^{-2}$, for a two-body binary system.  We discussed experiments which have
already verified the spin-orbit effects.  In fact, given the latter, the
results for the spin-spin effects must conform to the requirement that the total
angular momentum (orbital + spin of body 1 + spin of body 2) is conserved.

\section*{NOTE ADDED IN PROOF}

Our result for spin-orbit precession in a 2-body system \cite{barker75} has been verified, to an accuracy of 13\%, by Breton \emph{et al.} \cite{breton08} for the precession of pulsar \emph{B} in the double binary system.  In addition, in \cite{feiler09}, we presented, in Table 2, a comparison of numerical results for a variety of one-body and two-body systems.

\end{document}